\documentclass[12pt,psfig]{iopart}
\begin{document}

\title[Density-functional theory of a lattice gas]
{Density-functional theory of a lattice-gas model \\
with vapour, liquid, and solid phases}

\author{
Santi Prestipino\dag
\footnote[3]{To whom correspondence should be addressed
(Santi.Prestipino@unime.it)} and Paolo V. Giaquinta\dag
}

\address{\dag\ Istituto Nazionale per la Fisica della Materia
(INFM), Unit\`a di Ricerca di Messina, Italy; and
Universit\`a degli Studi di Messina, Dipartimento di Fisica,
Contrada Papardo, 98166 Messina, Italy
}

\begin{abstract}
We use the classical version of the density-functional theory in
the weighted-density approximation to build up the entire phase
diagram and the interface structure of a two-dimensional lattice-gas
model which is known, from previous studies, to possess three stable
phases -- solid, liquid, and vapour. Following the common practice,
the attractive part of the potential is treated in a mean-field-like
fashion, although with different prescriptions for the solid and the
fluid phases. It turns out that the present theory, compared to similar
theories in the continuum, is of worse quality. Nevertheless, at
least a number of qualitative facts are reproduced correctly:
i) the existence of three phases; ii) the disappearance of the
liquid phase when the range of the attraction is progressively
reduced; and iii) the intrusion, just below the triple-point
temperature, of a liquid-like layer at the interface between
the coexisting solid and vapour phases.

\end{abstract}

\pacs{05.20.Jj, 05.70.-a, 61.20.Ne, 64.10.+h, 68.08.Bc}

\maketitle


\section{Introduction}

Nowadays, most of the theoretical studies of the phase behaviour of a
classical fluid are formulated in the language of the density-functional
theory (DFT)~\cite{Evans,reviews}. Within such a theory,
the crystalline solid is viewed to be like an inhomogeneous system
with a periodically modulated density profile $n(x)$, whose free
energy is obtained through the optimization of a density functional
$F[n]$ which is built upon the structural properties of the fluid.
In particular, a successful recipe for $F[n]$ (sometimes called the
Hohenberg-Kohn-Mermin (HKM) free energy) is a local mapping into the
free energy of a homogeneous fluid with a suitably chosen effective
density $\bar{n}(x)$, which is related in a non-local way to the real
density $n(x)$. At variance with $n(x)$, the smoothed density
$\bar{n}(x)$ is a slowly-varying function of the position. This
general method has been named the ``weighted-density approximation''
(WDA)~\cite{Tarazona1,Curtin1}.

This scheme has proved to be sufficient for purely hard-core systems
and soft-repulsive ones. However, it badly fails in the presence of
attractive interactions, where it may happen that the solid is being
mapped onto a fluid with a density falling into the condensation gap
where, actually, no homogeneous phase is present. Nor it can be of
help the fundamental-measure approach of Rosenfeld~\cite{Rosenfeld},
which was recently extended from hard spheres~\cite{Rosenfeld,Tarazona2}
to spherically repulsive interactions~\cite{Schmidt}, but not yet to
attractive fluids. In these cases, the only available method is
lowest-order perturbation theory~\cite{Evans}, which is tantamount to
split the HKM free energy into the sum of the density functional for
a reference system (usually, a hard-core system) and a remainder,
containing the pair-distribution function of the inhomogeneous
reference system. A sensible approximation for the latter would
allow to draw accurate coexistence lines for the system under
consideration. In the past, a scheme of this sort has been
successfully applied by a number of authors to the prototypical
case being represented by the truncated Lennard-Jones
fluid~\cite{Curtin2,Tang,Mederos,Ohnesorge,Sweatman}.

In the present paper, we prove that this method is effective, although with
less quantitative success, also for lattice problems. As a case-study,
we shall focus on a two-dimensional (2D) lattice-gas system which is
known~\cite{Prestipino} to possess three phases with the features of
vapour, liquid, and solid. It is argued in Ref.\,\cite{Prestipino}
that the existence in this model of a further liquid phase, besides
the gaseous one, is made possible by the relatively long range of the
interparticle attraction. We point out that the choice of a 2D (rather
than 3D) system is only aimed at simplifying the forthcoming analysis
of the interface problem. In this respect, a crucial test for our DFT
will be the prediction of surface melting, which is actually in the
possibility of the DFT, as is proved by the 3D continuum theory of
Ref.\,\cite{Ohnesorge}.

While the general DFT framework on a lattice ({\it i.e.}, minimum
principle for the generalized grand potential plus Ornstein-Zernike
relation) is already known~\cite{Nieswand,Reinel}, we are not aware
of any single example of the performance of the lattice DFT in the
WDA for a system with a realistic phase diagram. We believe that
testing the degree of sophistication of the lattice-DFT method in a
rich context, such as that provided by a three-phase lattice-gas model,
can be interesting on fairly general grounds, {\it e.g.} for weighing up
the superiority, if any, of the DFT over other available statistical
methods like the transfer-matrix~\cite{Prestipino} or the cluster
variational methods~\cite{An}.

This paper is arranged as follows: After an outline, in section 2,
of the main contents of the lattice DFT, we describe our system
and method in section 3, and present our results for the bulk
of the system in section 4. Next, in section 5, we analyse the
structure of the interface between two bulk phases, including a
demonstration of the phenomenon of surface melting. Further remarks
and a brief summary of the main results are given in the Conclusions.

\section{Lattice density-functional theory}

We first review the lattice analogue of the classical DFT, which was
first considered by Nieswand {\it et al.} in 1993~\cite{Nieswand}.
Like the parent theory in the continuum, the lattice DFT is meant
to provide a general framework for discussing the statistical
properties of particles living on a regular lattice, in the
presence of a site-dependent external potential or, even, of a
self-sustained spatial inhomogeneity (like the one which, in a
simple fluid, becomes manifest at freezing). If lattice sites are
allowed to be occupied by at most one particle, a general Hamiltonian
for our problem is $H+\sum_i\epsilon_ic_i$, with
\begin{equation}
H=\sum_{i<j}v(|i-j|)c_ic_j\,.
\label{eq01}
\end{equation}
We call $c_i=0,1$ the occupation number of site $i$ in the lattice,
while $v(|i-j|)$ and $\epsilon_i$ are the pair-interaction and the
external potential, respectively.

The grand-canonical partition function
\begin{equation}
\Xi=\sum_{\{c\}}\exp\left( \beta\sum_i(\mu-\epsilon_i)c_i\right)
\exp\left( -\beta H\right)
\label{eq02}
\end{equation}
(with $\beta$ and $\mu$ representing the inverse temperature and the
chemical potential, respectively) is a {\it lattice functional}, namely
a function of all components $\mu_i=\mu-\epsilon_i$ of a lattice field,
which we call the external field.
Then, it is a simple matter to show that the (number-) density field
is given by:
\begin{equation}
n_i\equiv\left< c_i\right> =\frac{1}{\Xi}\frac{\partial\Xi}{\partial
\beta\mu_i}=-\frac{\partial\Omega}{\partial\mu_i}\,,
\label{eq03}
\end{equation}
which is a functional of the external field (a temperature dependence
is also implied).
$\Omega=-\beta^{-1}\ln\Xi$ is the grand potential, also a functional of
$\{\mu_i\}$. At variance with the continuum case, the density field is a
{\it partial}, rather than a functional, derivative of $\Xi$.

The Hohenberg-Kohn-Mermin theorem, which holds also for lattice gases,
ensures that there is a one-to-one correspondence, within the space of
$\mu$-representable densities, between the density field and the external
field. Thanks to this theorem, the Legendre transform of $\Omega[\mu]$
with respect to its functional variable is a well-defined object, which
generalizes the concept of Helmholtz free energy to inhomogeneous situations:
\begin{equation}
F[n]=\left. \Omega[\mu]+\sum_i\mu_in_i\right| _{\mu_i=\mu_i[n]}\,.
\label{eq04}
\end{equation}
Another expression for $F[n]$ is obtained by considering
\begin{equation}
\pi[n]=\frac{1}{\Xi}\exp\left( \beta\sum_i\mu_i[n]c_i\right)
\exp\left( -\beta H\right)
\label{eq05}
\end{equation}
(with $\Xi$ evaluated at $\mu_i[n]$),
which represents the grand-canonical probability density for the
given $n$, {\it i.e.}, the one calculated for $\mu_i=\mu_i[n]$. It
follows that:
\begin{equation}
F[n]=\sum_{\{c\}}\pi[n]\left( H+\frac{1}{\beta}\ln\pi[n]\right) \,.
\label{eq06}
\end{equation}
In practice, the ideal-gas system ($H=0$) is the only lattice system
for which the computation of the HKM free energy can be carried out
explicitly, with the result:
\begin{equation}
\beta F^{\rm id}[n]=\sum_i\left[ n_i\ln n_i+(1-n_i)\ln(1-n_i)\right] \,.
\label{eq07}
\end{equation}
In the general case, $F[n]$ is written as the sum of the ideal term and an
excess contribution $F^{\rm exc}[n]$ which is to be approximated some way.

A further density functional can be defined from $F[n]$, which is a sort
of generalized grand potential:
\begin{equation}
\Omega_{\mu}[\rho]=F[\rho]-\sum_i\mu_i\rho_i=\sum_{\{c\}}\pi[\rho]\left(
H+\frac{1}{\beta}\ln\pi[\rho]-\sum_i\mu_ic_i\right) \,,
\label{eq08}
\end{equation}
where a different symbol, $\rho$, is used for the density field to
stress the fact that no relation is implied between $\rho$ and $\mu$,
which should thus be regarded as independent functional variables.
Instead, we reserve the symbol $n$ for the density field derived
from $\mu$.

$\Omega_{\mu}[\rho]$ is actually the inhomogeneous Gibbs-Bogoliubov
functional of the classical mean-field theory. Hence, a minimum
principle holds, saying that $\Omega_{\mu}[\rho]$ attains its minimum
value for a density profile which is precisely the one determined by
the given external field, namely $n$. Moreover, the minimum
$\Omega_{\mu}[n]$ is nothing but the grand potential $\Omega$ for
the given $\mu$. The necessary condition for the minimum reads:
\begin{equation}
\left. \frac{\partial\Omega_{\mu}}{\partial\rho_i}\right| _{\rho_i=n_i}=0\,.
\label{eq09}
\end{equation}

The minimum principle for $\Omega_{\mu}[\rho]$ is, besides the HKM
theorem, the basic tenet of the DFT, being at the heart of a broad
family of approximate theories of freezing~\cite{reviews}. Every
such a theory starts from a prescription for $F[n]$, which is then
used to determine an approximate grand potential for the system
from which the thermodynamic properties are deduced.

We conclude our general presentation of the lattice DFT with the
Ornstein-Zernike (OZ) relation. Taken
\begin{equation}
c^{(1)}_i[n]=-\beta\frac{\partial F^{\rm exc}}{\partial n_i}
\,\,\,\,\,{\rm and}\,\,\,\,\,
c^{(2)}_{ij}[n]=\frac{\partial c^{(1)}_i}{\partial n_j}
\label{eq10}
\end{equation}
to be the one- and two-point direct correlation function (DCF),
respectively, the formal solution of Eq.\,(\ref{eq09}) reads:
\begin{equation}
n_i=\frac{1}{1+\exp\left\{ -\beta\mu_i-c^{(1)}_i[n]\right\}}\,.
\label{eq11}
\end{equation}
Upon introducing the reduced pair distribution function (PDF),
\begin{equation}
g_{ij}=(1-\delta_{ij})\frac{\left< c_ic_j\right> }{n_in_j}\,,
\label{eq12}
\end{equation}
and a further function
\begin{equation}
C_{ij}=c^{(2)}_{ij}-\frac{\delta_{ij}}{1-n_i}\,,
\label{eq13}
\end{equation}
it can be shown~\cite{Nieswand} that the following relation follows
from Eq.\,(\ref{eq11}):
\begin{equation}
h_{ij}=C_{ij}+\sum_kC_{ik}n_kh_{kj}\,,
\label{eq14}
\end{equation}
which is the lattice OZ relation ($h_{ij}=g_{ij}-1$ is called the
total correlation function). Only a further relation between $g$
and $c^{(2)}$ would allow one to determine both functions.
The importance of $c^{(2)}$ is two-fold: on one side, its knowledge
permits to recover, through the OZ relation, the PDF profile.
On the other side, most of the popular DFT approximations use
an expression of $F[n]$ in terms of the DCF of the fluid.

A number of simplifications occur for a homogeneous system ({\it i.e.},
one with $\epsilon=0$). Owing to translational symmetry, the one-point
DCF is a constant, $c^{(1)}_i[n]=c_1(\rho)$, whereas the two-point DCF
is a function of $i-j$ only, that is $c^{(2)}_{ij}[n]=c_2(i-j,\rho)$.
Furthermore, Eq.\,(\ref{eq14}) can be Fourier-transformed~\cite{note}
to give $\tilde{h}_q=\sum_xh_x\exp(-{\rm i}q\cdot x)$ in terms of
$\tilde{C}_q$ as:
\begin{equation}
\tilde{h}_q=\frac{\tilde{C}_q}{1-\rho\tilde{C}_q}\,,
\label{eq15}
\end{equation}
$\rho$ being the constant value of the density.

In the next section, we describe a DFT aimed at reconstructing the
phase diagram of a realistic lattice gas, that is one with a phase
diagram containing, besides a solid phase, also {\it two} different
fluid phases, liquid and vapour. In order to speed up the discussion,
we have decided to confine most of the technicalities to a few
appendices. In particular, Appendix A illustrates the lattice
counterpart of two celebrated, yet simple, theories of freezing:
the Ramakrishnan-Yussouff (RY) theory and the Tarazona's WDA.
We suggest the reading of Appendix A before proceeding to the
next paragraph.

\section{Model and method}

We shall work with the t345 model of Ref.\,\cite{Prestipino}.
This is a triangular-lattice model with a hard core covering
first- and second-neighbor sites and a pair attraction ranging
from third- to fifth-neighbor sites (see Fig.\,1 of
Ref.\,\cite{Prestipino}). The strength of the attraction reduces
upon increasing the distance from a reference site: whence, a
triangular solid is stable at high density and low temperature,
with a maximum density of $\rho_{\rm max}=0.25$. Upon comparing
the solid and the vapour grand potentials, one can easily predict
the zero-temperature value of the chemical potential at coexistence
to be $\mu_{\rm c}(T=0)=3v_3$, $v_3<0$ being the pair-potential
value at contact. To be specific, we use hereafter the same $v$
values that were considered in Ref.\,\cite{Prestipino}, namely
$v_3=-1.5V,v_4=-1.2V$, and $v_5=-V$, with $V>0$. In that paper,
a combination of transfer-matrix calculations and Monte Carlo (MC)
simulations distinctly showed the existence of a narrow temperature
interval where the increase of $\mu$, starting from large negative
values, drives the system through a couple of sharp (first-order)
phase transitions, {\it i.e.}, vapour-liquid and liquid-solid, as
is also revealed by the $\mu$-evolution of the number-density
histogram at fixed temperature. For later convenience, two other
models are introduced: the t3 model, which is the same as t345 but
with $v_4=v_5=0$, and the t model, where also $v_3=0$ and only the
hard-core interaction is present. At variance with the t345 case,
the MC simulation supports the existence of a unique fluid phase
in both the t and t3 models.

The first step in a typical DFT calculation is the determination of
an accurate DCF for the homogeneous system. In fact, an approximate
$F[n]$ is usually built upon this function (see Appendix A). The fluid
DCF is the solution to the homogeneous OZ relation plus a closure.
For 3D hard spheres, the most celebrated closure of all is the
Percus-Yevick approximation (PYA), which allows an exact determination
of the DCF~\cite{Hansen}. For a lattice system, the mean spherical
approximation (MSA) is easier to implement than the PYA, since
it leads to a smaller set of unknown quantities.
As a matter of fact, serious convergence problems are encountered
when trying to solve numerically either the MSA or the PYA of the
t345 fluid. Instead, no such problems occur for the MSA of the t
or, even, the t3 model (see details in Appendix B), while the PYA
of the t3 model is still intractable. As a result, we see us forced to
treat the t345 model perturbatively, as we are going to see in a moment
(note that our derivation of the perturbation formula, Eq.\,(\ref{eq16})
below, will be different from that of Evans~\cite{Evans}).

Let us write $v(|i-j|)$ as $v_0(|i-j|)+\Delta v(|i-j|)$, where $v_0$
describes a reference system, say the t model, and $\Delta v$ is a
remainder. We shall prove that, using a 0 subscript for quantities
pertaining to the t model, one has, at the lowest order in $\beta$:
\begin{equation}
F[n]=F_0[n]+\sum_{i<j}\Delta v(|i-j|)\left< c_ic_j\right> _0\,.
\label{eq16}
\end{equation}

Let $v_{\lambda}=v_0+\lambda\Delta v$ be a linear path between $v_0$
and $v$, with $0\leq\lambda\leq 1$. Accordingly, we define
$H_{\lambda}=H_0+\lambda\Delta H$. Let $\pi_{\lambda}[n]$ be the
grand-canonical probability density of $H_{\lambda}$ under the
condition that the external field takes precisely that value,
$\{\mu_{\lambda i}[n]\}$, which produces a density of $n$.
Next, we define:
\begin{equation}
F_{\lambda}[n]=\sum_{\{c\}}\pi_{\lambda}[n]\left( H_{\lambda}+\frac{1}{\beta}
\ln\pi_{\lambda}[n]\right)
\label{eq17}
\end{equation}
to be the HKM free energy relative to $H_{\lambda}$.

In the same spirit of classical Zwanzig perturbation theory~\cite{Hansen},
we derive an approximate expression for $F[n]$ starting from the exact
formula:
\begin{equation}
F[n]=F_0[n]+\int_0^1{\rm d}\lambda\,\frac{\partial F_{\lambda}[n]}
{\partial\lambda}\,.
\label{eq18}
\end{equation}
A rather lengthy calculation gives:
\begin{equation}
\fl \frac{\partial F_{\lambda}[n]}{\partial\lambda}=
\left< \Delta H\right> _{\lambda}+\beta\left[
\left< B\right> _{\lambda}\left< \Delta H\right> _{\lambda}-
\left< B\Delta H\right> _{\lambda}\right] -\beta\left[
\left< A\right> _{\lambda}\left< B\right> _{\lambda}-
\left< AB\right> _{\lambda}\right] \,,
\label{eq19}
\end{equation}
where $\left< \ldots\right> _{\lambda}$ is an average over
$\pi_{\lambda}[n]$ and
\begin{equation}
A=\sum_i\frac{\partial\mu_{\lambda i}[n]}{\partial\lambda}c_i
\,\,\,\,\,{\rm and}\,\,\,\,\,
B=\sum_i\mu_{\lambda i}[n]c_i\,.
\label{eq20}
\end{equation}
Considering that $\mu_{\lambda i}[n]$ is unknown,
some assumption must be made in order to obtain $F[n]$.
In particular, if $v_0$ is a hard-core interaction, the r.h.s.
of Eq.\,(\ref{eq19}) reduces, at the lowest order in $\beta$, to
$\left< \Delta H\right> _0$, yielding eventually Eq.\,(\ref{eq16}).

We note that, in Eq.\,(\ref{eq16}), $\left< c_ic_j\right> _0=
n_in_jg_{0,ij}[n]$ contains the exact, yet unknown, reduced PDF
of the inhomogeneous t model. Hence, the above equation is useless
unless one finds a careful prescription for $g_{0,ij}$, which could
be possibly {\it different} for the fluid and solid phases. Before
discussing this point further, we go back for a moment to the
reference system.

After obtaining the DCF of the homogeneous t system, we use
Eq.\,(\ref{a03}) to calculate the fluid excess free energy per
particle $f^{\rm exc}(\rho)$. This quantity, which is a monotonously
increasing function of the density, ceases to be defined at
$\rho\simeq 0.21$, beyond which no MSA solution is actually found.
However, this density is too small for allowing a description
(within the WDA) of the very dense solid. Hence, the problem arises
as to what criterion should be used in order to extrapolate
$f^{\rm exc}(\rho)$ beyond that limit. This problem is discussed
in Appendix B, where two different solutions are proposed. Here,
suffice it to say that there exists a method to prolong the
definition of $f^{\rm exc}(\rho)$ insofar as needed, with all
regularity requirements fully met.

We have sketched in Appendix B the details of a simple DFT (the RY
theory~\cite{Haymet}) for the freezing of the t model. However, in
order to have a good description of the reference system, we have
tried to do better than the RY theory. In fact, the stability of
the liquid phase is a matter of a delicate balance between energy
and entropy; hence, an accurate representation of the solid free
energy is an obvious necessity in all cases where a liquid phase
is expected. Leaving aside Rosenfeld's fundamental-measure theory,
whose extension to lattice fluids is not immediate (see, however,
the recent contribution \cite{Lafuente}), we have applied the
lattice counterpart of the WDA in the version implemented by
Tarazona~\cite{Tarazona1}, which gives rather good results for the
hard-sphere system. This theory is reviewed in the Appendices A
(general) and C (t model). Here, we provide just a few details on
the method.

The hypothesis underlying any WDA is an approximation of the excess
free energy of the system as $\sum_in_if^{\rm exc}(\bar{n}_i)$, where
the weighted density $\bar{n}_i$ is a non-local functional of the
density field, given implicitly by $\bar{n}_i=\sum_jn_jw(i-j,\bar{n}_i)$.
In turn, the weight function $w(i-j,\rho)$ is such that both the
density and the DCF of the fluid are recovered in the homogeneous limit.
In the Tarazona's WDA, the further assumption is made that the weight
function is a second-order polynomial in the density $\rho$. We thus
have a well-defined algorithm to build up the excess free energy and,
eventually, the density functional that is used to trace the conditions
for fluid-solid coexistence.

Once the free energy of the reference system is given, we are left
with the problem of incorporating the attraction $\Delta v$ into
the density functional of the t system using Eq.\,(\ref{eq16}). We
shall distinguish between the fluid and the solid, although this way
the HKM functional will only approximately be the same for all phases
(this will have some harmful consequences for the interface structure,
see section 5). While we obviously use for the fluid the reduced PDF
of the homogeneous t system, as far as the solid is concerned we shall
make the (apparently bad) approximation $g_{0,ij}[n]=1$ outside the core,
which is the same assumption of the mean-field approximation (MFA).
In fact, we agree with Mederos {\it et al.}~\cite{Mederos} that
the PDF of the low-temperature solid is {\it trivial}, since all the
structure (which in a fluid is accounted for by the reduced PDF) is
already present at the level of $n_i$ itself~\cite{note2}. This is
rather obvious at $T=0$, where $g_{ij}=1$ at the typical distances
of the perfect solid, while being undefined elsewhere. For small, but
non-zero temperatures, a quasi-random (ideal-gas-like) distribution of
interstitials and vacancies would extend the result $g_{ij}\simeq 1$
to all distances outside the core region.

A more refined approximation for the attractive interaction would be
that of Ref.\,\cite{Mederos}. This theory uses the same prescription
for the solid and the fluid, based on the use of the compressibility
sum rule. However, the implementation of this method is also very
difficult and much more involved than ours. In particular, the two
algorithms for minimization that are described in Appendix B do both
require the numerical evaluation of the density derivatives of the DFT
functional, which is indeed a very difficult task to accomplish if the
recipe of Ref.\,\cite{Mederos} is followed. For the sake of truth, we
have also attempted to use the method of Ref.\,\cite{Curtin2}. This
relies on two approximations: i) the use of Eq.\,(\ref{eq16}) for the
t345 {\it fluid}; and ii) the decomposition of the DCF of the
inhomogeneous t345 system as the sum of the analogous function for the
t system and a remainder $\Delta c_2(i-j,\rho)$, assumed to be zero
inside the core and MSA-like outside this region. As a matter of fact,
we found no stable liquid phase by this method.

Going back to our theory, we write the difference in grand potential
between the triangular solid (whose density field can be parametrized
by means of two numbers only, see below) and the fluid with equal $T$
and $\mu$ as the minimum of:
\begin{eqnarray}
\fl \Delta\Omega(n_A,n_B)=\Delta\Omega^{{\rm (t)}}(n_A,n_B)
+\frac{N}{4}k_BT\left\{ 3\beta v_3n_A^2+(12\beta v_4+6\beta v_5)n_An_B\right.
\nonumber \\
\lo+(9\beta v_3+12\beta v_4+6\beta v_5)n_B^2
-2\rho^2\sum_{n=3}^5z_n\beta v_ng_0(n,\rho)
\nonumber \\
\lo-\left. \left( \rho\sum_{n=3}^5z_n\beta v_ng_0(n,\rho)+\frac{\rho^2}{2}
\sum_{n=3}^5z_n\beta v_n\frac{{\rm d}g_0(n,\rho)}{{\rm d}\rho}\right)
(n_A+3n_B-4\rho)\right\} \,.
\label{eq21}
\end{eqnarray}
Note that, in the above equation: i) $n_A$ and $n_B$ are the number
densities in the sublattices $A$ and $B$ of occupied and unoccupied
sites, respectively (see Appendix B); ii) $\Delta\Omega^{({\rm t})}$
is the functional for the t model, defined at Eq.\,(\ref{c01});
iii) $z_n$ is the coordination number
for the $n$-th shell, that is $z_3=6,z_4=12$, and $z_5=6$; and iv)
$g_0(n,\rho)$ is the value taken by the reduced PDF of the t fluid
at the $n$-th-neighbour separation. Apart from a different density
dependence in $\Delta\Omega$, the machinery needed for calculating
the weighted densities $\bar{n}_A$ and $\bar{n}_B$ (and their
derivatives) from the densities $n_A$ and $n_B$ remains the
same as for the t model, illustrated in Appendix C.

The equations for $n_A$ and $n_B$ are then (see, for comparison,
Eqs.\,(\ref{c02})):
\begin{eqnarray}
\fl n_A^{-1}=1+\frac{1-\rho}{\rho}
\exp\left\{ c_1(\rho)+\beta f^{\rm exc}(\bar{n}_A)+n_A\beta
f^{\rm exc\,\prime}(\bar{n}_A)\frac{\partial\bar{n}_A}{\partial n_A}+3n_B\beta
f^{\rm exc\,\prime}(\bar{n}_B)\frac{\partial\bar{n}_B}{\partial n_A}\right.
\nonumber \\
\lo- \left[ \rho\sum_nz_n\beta v_ng_0(n,\rho)+\frac{\rho^2}{2}
\sum_nz_n\beta v_n\frac{{\rm d}g_0(n,\rho)}{{\rm d}\rho}\right]
\nonumber \\
\lo+ \left. 6\beta v_3n_A+(12\beta v_4+6\beta v_5)n_B\right\} \,;
\nonumber \\
\fl n_B^{-1}=1+\frac{1-\rho}{\rho}
\exp\left\{ c_1(\rho)+\beta f^{\rm exc}(\bar{n}_B)+\frac{1}{3}n_A\beta
f^{\rm exc\,\prime}(\bar{n}_A)\frac{\partial\bar{n}_A}{\partial n_B}+n_B\beta
f^{\rm exc\,\prime}(\bar{n}_B)\frac{\partial\bar{n}_B}{\partial n_B}\right.
\nonumber \\
\lo- \left[ \rho\sum_nz_n\beta v_ng_0(n,\rho)+\frac{\rho^2}{2}
\sum_nz_n\beta v_n\frac{{\rm d}g_0(n,\rho)}{{\rm d}\rho}\right]
\nonumber \\
\lo+ \left. (4\beta v_4+2\beta v_5)n_A
+(6\beta v_3+8\beta v_4+4\beta v_5)n_B\right\} \,.
\label{eq22}
\end{eqnarray}
In Appendix B, we have outlined two different numerical algorithms for
solving the minimum problem for a functional of the kind of (\ref{eq21}).

We conclude our survey of the method with a few words about the
liquid-vapour phase transition in the t345 model. The generalized
grand potential of the homogeneous t345 system is
$\Omega_{\mu}(\rho)=F(\rho)-N\mu\rho\equiv N(a(\rho)-\mu\rho)$, where
\begin{equation}
\fl \beta a(\rho)=\rho\ln\rho+(1-\rho)\ln(1-\rho)+\rho\beta f^{\rm exc}(\rho)+
\frac{1}{2}\rho^2\sum_{n=3}^5z_n\beta v_ng_0(n,\rho)\,.
\label{eq23}
\end{equation}
At low enough temperature, there exists an
interval of $\mu$ values where the minima of $\Omega_{\mu}(\rho)$
are in fact two, corresponding to the competing vapour and liquid phases
(while the deeper minimum yields the physical solution, the other
is associated with a metastable state). In particular, if we call
$\rho_{\rm v}$ and $\rho_{\rm l}$ the related densities, the
coexistence of the two phases occurs when the minima are equal:
\begin{equation}
\Omega_{\mu}(T,\rho_{\rm v})=\Omega_{\mu}(T,\rho_{\rm l})
\,\,\,\,\,{\rm and}\,\,\,\,\,
\Omega_{\mu}^{\prime}(T,\rho_{\rm v})=
\Omega_{\mu}^{\prime}(T,\rho_{\rm l})=0\,.
\label{eq24}
\end{equation}
The above equations are easily identified with the thermodynamic
conditions for phase coexistence, {\it i.e.}, equal values of
$T,P$ (the pressure), and $\mu$ for the two phases. This will
automatically give rise to the Maxwell construction for the
pressure and will also provide the right position where to cut
the non-monotonous profile $a^{\prime}(\rho)$ of the chemical
potential as a function of the density.

\section{DFT results: bulk}

In this section, we present the results that we have obtained for
the bulk properties of the t345 model by the lattice-DFT method
outlined in the previous section. In order to check them,
we have resorted to the MC simulation. In a typical
grand-canonical MC experiment, a lattice-gas system is driven to
equilibrium by a series of moves (creation or annihilation of one
particle at a time), which are designed in such a way as to
satisfy detailed balance (for more details, the reader is referred
to \cite{Prestipino}). In particular, a first-order transition is
located at those values of $T$ and $\mu$ where the number-density
hystogram of a large system sample shows two peaks of equal height,
signalling that two distinct phases are equally stable.

We first review our results for the t model. We have formulated two
different DFTs for the freezing of this model, {\it i.e.}, the RY
theory and the WDA of Tarazona. While the results of the former are
discussed in Appendix B, an outline of the latter can be found in
Appendix C. Both theories rely on a MSA description for the fluid.
Within the WDA theory, the densities of the coexisting fluid and
solid are found to be $\rho_{\rm f}=0.1335$ and $\rho_{\rm s}=0.1686$.
Whence, a considerably larger density jump is predicted at the transition
than given by the RY theory. Anyway, these numbers are still very far
from those obtained by MC, {\it i.e.}, $\rho_{\rm f}=0.172(1)$ and
$\rho_{\rm s}=0.188(1)$, indicating that the instability of the t
fluid against the solid is strongly anticipated in the WDA. As for
the chemical-potential value at coexistence, the agreement with MC
is also poor: while the WDA gives (through Eqs.\,(\ref{a04})
and (\ref{a05})) $\mu_{\rm c}=1.2655\,V$, MC yields instead
$\mu_{\rm c}=1.725(5)\,V$.

In Fig.\,1, the local and the weighted density of the t model are
separately plotted for the two sublattices as a function of the chemical
potential. In particular, the weighted density takes its larger value
in the interstitial region, that is in the $B$ sublattice. This is a
counterintuitive effect which, however, is not peculiar to the lattice,
being also found in the continuum (see, for instance, Ref.\,\cite{Curtin1}).

Moving to the t345 model, we have first checked the existence of
two distinct fluid phases at low temperature. The liquid-vapour
coexistence line is drawn by solving Eqs.\,(\ref{eq23}) and
(\ref{eq24}) (see the following Fig.\,2). This gives a critical
point at $t_{\rm cr}=1.27(1)$ and $\rho_{\rm cr}=0.079(2)$
(hereafter, reduced units $t=kT/V$ are used for the temperature).
Also shown in Fig.\,2 is the coexistence line as predicted by the
MFA. The latter also uses Eq.\,(\ref{eq23}), but with a 1 in place
of $g_0(n,\rho)$.

Finally, we have minimized the density functional (\ref{eq21}) in
order to obtain the freezing and melting lines of the t345 model.
It is right at this point that the choice between E1 and E2 (for
extrapolating $f^{\rm exc}(\rho)$ beyond $\rho=0.21$, see Appendix
B) becomes crucial. In fact, while the solid phase never becomes
stable -- below a certain temperature -- if E1 is adopted, we never
run into troubles if extrapolation E2 is used. Anyway, E2 gives
practically the same results as E1 at high temperature.

The complete DFT phase diagram of the t345 model is plotted in
Fig.\,2 (open circles), together with the results of the MFA
(crosses) and MC simulation (asterisks). To our delight, a triple
point eventually shows up in the t345 phase diagram, at $t_{\rm
tr}=1.145(5)$ and $\rho_{\rm tr}=0.122(1)$, as long as different
forms of the perturbation part are used in the HKM functional for
the solid and for the fluid. In other words, the use of $g_0$ for
the description of the reduced PDF of the fluid turns out to be
essential for obtaining a liquid region in the phase diagram.
The liquid phase is unstable if the MFA is used also for the
fluid. However, the agreement of our DFT with the MC results is
mainly qualitative: the exact coordinates of the triple point are
very different, $t_{\rm tr}^{\rm MC}=0.87(1)$ and $\rho_{\rm
tr}^{\rm MC}=0.191(1)$; only the ratio of $t_{\rm tr}$ to $t_{\rm
cr}$ is similar.

As a matter of example, we have plotted in Fig.\,3 the $\mu$-evolution
of the DFT number density at $t=1.2$, upon going across the two phase
transitions. Finally, Fig.\,4 shows the DFT phase diagram of the t345
model in the $T$-$\mu$ plane, where we recognize the typical fork with
two teeth of different length. In the same picture, the MC data points
of Fig.\,2 are also reported for comparison.

We have studied the t3 model with the same DFT described above in
order to check the internal consistency of our method.
For this model, Eq.\,(\ref{eq23}) with $v_4=v_5=0$ does never produce
two distinct fluid phases, and the freezing and melting lines are
similar to those found by the simpler RY theory (see Fig.\,5).
This result can be rationalized as follows: in the t345 model,
the existence of attractive sites at the ``interstitial'' distances
$r_4$ and $r_5$ causes the upper stability threshold of the fluid
phase to move up in density with respect
to the t3 model, thus contributing to unveil the triple point.
This effect is missing in the t3 model, which thus fails to
become a liquid. The conclusion, in perfect agreement with MC,
is that no liquid phase is present in the t3 model.

Finally, we make a comment on the possible causes of the quantitative
failure of our DFT for the t345 model. On one side, one generally
expects that mean-field theory works well in 3D, less in 2D. One
should also not forget that the perturbation formula (\ref{eq16})
is a high-temperature approximation and that, at variance with the
continuum case, there is no Barker-Henderson criterion which can be
called for optimizing the hard-core diameter of the reference system.
On the other hand, also the low quality of the MSA for the reference
t fluid is partly responsible for the wrong position of the freezing
and melting lines. To overcome this problem, we have made an attempt
of replacing the MSA with the hypernetted-chain approximation (HNCA)
as a closure for the OZ relation of the homogeneous t system.
For this model, the HNCA assumes:
i) $h(0)=h(1)=h(2)=-1$ (here, the argument is the shell number);
ii) $C(i-j,\rho)=h(i-j,\rho)-\ln\left[ 1+h(i-j,\rho)\right] $, outside
the core. In practice, we should also assume that $C$ and $h$ are
exactly zero beyond a certain distance, and we have chosen this to
be the distance of the 38th neighbors ({\it i.e.}, $6\sqrt{3}$).
The solution method is iterative: at a given $\rho$, we make an
initial estimate of
$C(0),C(1),C(2)$ and $h(3),h(4),\ldots$, which are then updated
using the inverse of the Fourier transform (\ref{eq15}).
Unfortunately, however, this works only up to $\rho=0.11$,
which is too small a density for allowing us to build
an accurate reference-fluid free energy. Just in order to
appreciate the difference between the two OZ closures, we have
plotted in Fig.\,6 the profiles, for $\rho=0.1$, of the reduced
PDF of the t model as given by the MSA and by the HNCA, respectively.
The comparison with the ``exact'' MC profile at the same density
reveals the superiority of the MSA over the HNCA, which overestimates
the structure of the PDF. However, a good fluid structure is not
necessarily accompanied by good thermodynamic properties, and this is
actually the case of the MSA of the t model.

\section{DFT results: interfaces}

Now that we have an accurate density functional for the bulk of
the t345 system, we move on to consider the structure of the
interface between two coexisting bulk phases. Many similar
calculations have been carried out in the past (see, for instance,
Refs.\,\cite{Curtin3,Ohnesorge,Reinel}) and, in fact, the
development of more and more careful DFT-based microscopic
descriptions of the density profile across an interface has been
historically a recurrent {\it leitmotiv}~\cite{reviews}.

Here, two cases are analysed which will deserve a rather different
treatment: the liquid-vapour interface, {\it i.e.}, the interface
between two homogeneous phases, and the solid-fluid interface,
which instead separates a broken-symmetry phase from a homogeneous
one.

\subsection{Liquid-vapour coexistence}

As a first example, we have studied the interface between the
coexisting liquid and vapour phases of the t345 model. This
interface is assumed to lie perpendicularly to the $y$ direction.
Horizontal layers are labelled with an integer $\lambda$, which
is taken to be zero at the ``centre'' of the interface.
Since both phases are homogeneous, the density will be uniform
along the $x$ direction, its value being a constant,
$\rho_{\lambda}$, for all sites $i$ of the $\lambda$-th layer.
Let $\rho_{\rm l}$ and $\rho_{\rm v}$ be the densities of the
coexisting phases at a given temperature $T$. Then, the common
value $\mu$ of the chemical potential is
$a^{\prime}(\rho_{\rm v})=a^{\prime}(\rho_{\rm l})$ (with
$a(\rho)$ defined at Eq.\,(\ref{eq23})). For these $T$ and $\mu$,
the grand potential per site of the bulk vapour or liquid is
$a(\rho_{\rm v})-\mu\rho_{\rm v}=a(\rho_{\rm l})-\mu\rho_{\rm l}$.
Given that, the generalized grand potential of the inhomogeneous
system is:
\begin{eqnarray}
\fl \beta\Omega_{\mu}[\rho]=N_x\sum_{\lambda}\left[
\rho_{\lambda}\ln\rho_{\lambda}+(1-\rho_{\lambda})\ln(1-\rho_{\lambda})+
\rho_{\lambda}\beta f^{\rm exc}(\rho_{\lambda})\right]
\nonumber \\
\lo+ \frac{1}{2}N_x\sum_{\lambda}\rho_{\lambda}\sum_{j|i\in\lambda}n_j
\beta\Delta v(|i-j|)g_0\left( i-j,\frac{\rho_{\lambda}+n_j}{2}\right)
-N_x\beta\mu\sum_{\lambda}\rho_{\lambda}\,,
\label{eq25}
\end{eqnarray}
a functional of $\{\rho_{\lambda}\}$ being subject to the conditions
$\rho_{\lambda}\rightarrow\rho_{\rm l}$ for $\lambda\rightarrow -\infty$ and
$\rho_{\lambda}\rightarrow\rho_{\rm v}$ for $\lambda\rightarrow +\infty$.
As is usual practice~\cite{Evans}, the $g_0$ function of the inhomogeneous
t system at the $i-j$ lattice separation is represented by the fluid PDF
as calculated for a density which is the arithmetic mean of the local
densities in $i$ and $j$. Finally, the grand-potential excess per surface
particle due to the interface can be estimated as $\sigma(T)=\min_{n}
\Sigma[n]$, where:
\begin{equation}
\Sigma[n]=\frac{2\beta}{N_x}\left\{ \Omega_{\mu}[n]
-\Omega_{\mu}(\rho_{\rm v})\right\} \,.
\label{eq26}
\end{equation}

The calculation of $\sigma$, which is nothing but the surface
tension of the interface under consideration, proceeds in two
steps: one first optimizes a simple exp {\it ansatz}~\cite{Reinel}
and then refines the calculation via an unconstrained minimization
that is accomplished in a way analogous to that followed for the bulk.

We have chosen, for a demonstration, a temperature of $t=1.15$,
which is slightly above the triple-point temperature. For this
case, the shape of the liquid-vapour interface is plotted in
Fig.\,7. In this picture, the dotted curve represents the best
exp profile, while the continuous line is our
final optimization. The surface tension is thus found to be
$\sigma=0.0145$. By looking at Fig.\,7, it appears that the
deviation of the density profile from the exponential law is
actually minute.

\subsection{Solid-vapour coexistence}

We have first analysed the structure of the solid-fluid interface
in the t3 lattice gas by the RY theory, as built over the MSA DCF.
To be specific, we consider a linear interface running along $x$.
Such an interface breaks the translational
symmetry along $y$, thus causing the sublattice densities to vary
with $y$. Only very far from the interface, the densities
recover the bulk values, being those of the solid, say, far below
the interface and those of the coexisting fluid far above. The
horizontal layers are labelled with an integer index, $\lambda$,
which increases upon moving from the solid to the fluid region,
being zero at the interface. We choose {\it e.g.} odd $\lambda$
values for those layers where particles are hosted in the $T=0$
solid. At variance with the bulk case, we must distinguish {\it
three} sublattices since we generally expect different density
values at the interstitial sites pertaining to the even and to the
odd layers. We call $C$ the sublattice formed by the interstitial
sites in the odd layers, and $B$ the other. Finally, $A$ is the
triangular sublattice which is occupied in the $T=0$ solid. We
note that a $C$ site has two adjacent $A$ sites on the same layer.
Conversely, the two closest $A$ sites of a $B$ site stay on the
(odd) layers which are respectively below and above the (even)
layer which the $B$ site belongs to.

In the RY theory, the sublattice densities are drawn from
Eq.\,(\ref{eq11}) with $\mu_i=\mu$ (given by Eq.\,(\ref{a05}))
and a linear density-functionality is assumed for the one-point DCF:
\begin{equation}
c^{(1)}_i[n]=c_1(\rho)+\sum_jc_2(i-j,\rho)(n_j-\rho)\,,
\label{eq27}
\end{equation}
where $i$ can belong to $A,B$, or $C$. In particular, for the odd
values of $\lambda$ we have:
\begin{eqnarray}
\fl c^{(1)}(A,\lambda)=c_1(\rho)+c_2(0,\rho)(n_{A,\lambda}-\rho)+2c_2(1,\rho)
\left( n_{B,\lambda-1}+n_{C,\lambda}+n_{B,\lambda+1}-3\rho\right)
\nonumber \\
\lo+ c_2(2,\rho)\left( n_{C,\lambda-2}+2n_{B,\lambda-1}+2n_{B,\lambda+1}+
n_{C,\lambda+2}-6\rho\right)
\nonumber \\
\lo+ 2c_2(3,\rho)\left( n_{A,\lambda-2}+n_{A,\lambda}+n_{A,\lambda+2}-
3\rho\right) \,;
\nonumber \\
\fl c^{(1)}(C,\lambda)=c_1(\rho)+c_2(0,\rho)(n_{C,\lambda}-\rho)+2c_2(1,\rho)
\left( n_{B,\lambda-1}+n_{A,\lambda}+n_{B,\lambda+1}-3\rho\right)
\nonumber \\
\lo+ c_2(2,\rho)\left( n_{A,\lambda-2}+2n_{B,\lambda-1}+2n_{B,\lambda+1}+
n_{A,\lambda+2}-6\rho\right)
\nonumber \\
\lo+ 2c_2(3,\rho)\left( n_{C,\lambda-2}+n_{C,\lambda}+n_{C,\lambda+2}-
3\rho\right) \,,
\label{eq28}
\end{eqnarray}
while, for the even values of $\lambda$:
\begin{eqnarray}
\fl c^{(1)}(B,\lambda)=c_1(\rho)+c_2(0,\rho)(n_{B,\lambda}-\rho)+c_2(1,\rho)
\left( n_{A,\lambda-1}+n_{C,\lambda-1}+2n_{B,\lambda}+n_{A,\lambda+1}+
n_{C,\lambda+1}-6\rho\right)
\nonumber \\
\lo+ c_2(2,\rho)\left( n_{B,\lambda-2}+n_{A,\lambda-1}+n_{C,\lambda-1}+
n_{A,\lambda+1}+n_{C,\lambda+1}+n_{B,\lambda+2}-6\rho\right)
\nonumber \\
\lo+ 2c_2(3,\rho)\left( n_{B,\lambda-2}+n_{B,\lambda}+n_{B,\lambda+2}-
3\rho\right) \,.
\label{eq29}
\end{eqnarray}
Next, the RY density functional is derived from Eq.\,(\ref{a06}),
where it must be noted that
\begin{eqnarray}
\fl \sum_{i,j}c_2(i-j,\rho)(n_i-\rho)(n_j-\rho)=
\sum_i(n_i-\rho)\left\{ c^{(1)}_i[n]-c_1(\rho)\right\}
\nonumber \\
\lo= \frac{N_x}{2}\sum_{\lambda\,\,{\rm odd}}(n_{A,\lambda}-\rho)\left\{
c^{(1)}(A,\lambda)-c_1(\rho)\right\}+
N_x\sum_{\lambda\,\,{\rm even}}(n_{B,\lambda}-\rho)\left\{
c^{(1)}(B,\lambda)-c_1(\rho)\right\}
\nonumber \\
\lo+ \frac{N_x}{2}\sum_{\lambda\,\,{\rm odd}}(n_{C,\lambda}-\rho)\left\{
c^{(1)}(C,\lambda)-c_1(\rho)\right\} \,.
\label{eq30}
\end{eqnarray}
As is clear, the final expression of the functional (\ref{a06}) is
rather cumbersome and, therefore, we do not specify it here. Hence,
we directly move to the numerical results.

We have considered just one temperature value, $t=1.8036$.
At this temperature, the fluid and solid coexistence densities are
$\rho_{\rm f}=0.1000$ and $\rho_{\rm s}=0.1695$, respectively. Our
slab consisted of 61 layers, from $\lambda=-30$ to $\lambda=30$
(at the boundaries, we have set the sublattice densities fixed to
the solid values for $\lambda<-30$ and to the fluid value for
$\lambda>30$). To optimize the interface shape, we proceed
in two steps: first, we attempt a rough optimization by the simple
one-parameter {\it ansatz}~\cite{Reinel}:
\begin{eqnarray}
n_{A,\lambda} &=& \rho+\frac{n_A-\rho}{1+\exp(\lambda/l)}\,;
\,\,\,(\lambda\,\,{\rm odd})
\nonumber \\
n_{B,\lambda} &=& \rho+\frac{n_B-\rho}{1+\exp(\lambda/l)}\,;
\,\,\,(\lambda\,\,{\rm even})
\nonumber \\
n_{C,\lambda} &=& \rho+\frac{n_B-\rho}{1+\exp(\lambda/l)}\,.
\,\,\,(\lambda\,\,{\rm odd})
\label{eq31}
\end{eqnarray}
The parameter $l$ is chosen in such a way as to make (\ref{a06})
as low as possible. With that profile as a starting point,
we run an iterative procedure, similar to that used for the bulk,
for the unconstrained minimization of
$\Omega_{\mu}[\rho]-\Omega_{\mu}(\rho_{\rm f})$. In the end, we
get the density profile shown in Fig.\,8 (top). At this temperature,
the surface tension, given by Eq.\,(\ref{eq26}), takes the value
$\sigma=0.0740(1)$.

Next, we move to the t model, as described by the WDA theory outlined
in Appendix C. From Eq.\,(\ref{a05}), we obtain the following expression
for $\Sigma[n]$:
\begin{eqnarray}
\fl \Sigma[n]=\sum_{\lambda\,\,{\rm odd}}\left[
n_{A,\lambda}\ln\frac{n_{A,\lambda}}{\rho_{\rm v}}+
(1-n_{A,\lambda})\ln\frac{1-n_{A,\lambda}}{1-\rho_{\rm v}}+
n_{C,\lambda}\ln\frac{n_{C,\lambda}}{\rho_{\rm v}}+
(1-n_{C,\lambda})\ln\frac{1-n_{C,\lambda}}{1-\rho_{\rm v}}\right]
\nonumber \\
\lo+ 2\sum_{\lambda\,\,{\rm even}}\left[
n_{B,\lambda}\ln\frac{n_{B,\lambda}}{\rho_{\rm v}}+
(1-n_{B,\lambda})\ln\frac{1-n_{B,\lambda}}{1-\rho_{\rm v}}\right]
+c_1(\rho_{\rm v})\sum_{\lambda\,\,{\rm odd}}
(n_{A,\lambda}+n_{C,\lambda}-2\rho_{\rm v})
\nonumber \\
\lo+ 2c_1(\rho_{\rm v})\sum_{\lambda\,\,{\rm even}}(n_{B,\lambda}-\rho_{\rm v})
+\sum_{\lambda\,\,{\rm odd}}\left[
n_{A,\lambda}\beta f^{\rm exc}(\bar{n}_{A,\lambda})+
n_{C,\lambda}\beta f^{\rm exc}(\bar{n}_{C,\lambda})
-2\rho_{\rm v}\beta f^{\rm exc}(\rho_{\rm v})\right]
\nonumber \\
\lo+ 2\sum_{\lambda\,\,{\rm even}}\left[
n_{B,\lambda}\beta f^{\rm exc}(\bar{n}_{B,\lambda})
-\rho_{\rm v}\beta f^{\rm exc}(\rho_{\rm v})\right] \,.
\label{eq32}
\end{eqnarray}
For $\lambda$ {\it e.g.} even, the only weighted density that
matters is $\bar{n}_{B,\lambda}$, which is defined in terms of all
densities as:
\begin{equation}
\bar{n}_{B,\lambda}=\sum_jn_jw(i-j,\bar{n}_{B,\lambda})\,,
\label{eq33}
\end{equation}
where $i$ is any particular site on the $\lambda$-th layer. For
odd values of $\lambda$, one can analogously define
$\bar{n}_{A,\lambda}$ and $\bar{n}_{C,\lambda}$. If we adopt the
WDA method of Tarazona, then a result similar to Eq.\,(\ref{a15})
is obtained, giving $\bar{n}_{B,\lambda}$ in terms of the
auxiliary quantities
\begin{equation}
\bar{n}_{kB,\lambda}=\sum_jn_jw_k(i-j)\,\,\,\,\,\,
({\rm with }\,\,k=0,1,2)\,.
\label{eq34}
\end{equation}
The explicit expression of (\ref{eq34}) obviously requires the
careful consideration of lattice sites $j$ lying progressively
farther from the reference site $i$. As noted in Appendix B, a sum
like (\ref{eq34}) should in practice be truncated after a certain
value of $|i-j|$, and we have chosen this to be the distance of
the 20th neighbors. Even so, the final formula takes too many
lines to be specified here, and is therefore omitted.

The actual minimization of $\Sigma[n]$ proceeds in a way analogous to
the bulk case, described in Appendix C. However, the formulae for the
density derivatives of $\bar{n}_{A,\lambda},\bar{n}_{B,\lambda}$, and
$\bar{n}_{C,\lambda}$ are much more involved for the surface than for
the bulk case. The outcome for the density profile across the interface
is shown in Fig.\,8 (bottom). Its shape is very similar to that of the
t3 model, but the surface tension is much larger, our best result being
$\sigma=0.3182$.

We have finally considered the solid-vapour interface in the t345
model. In particular, we are interested in temperature values that
are just below the triple-point temperature. In such conditions,
and as long as surface melting occurs, a thin liquid-like film
appears at the interface between the solid and the vapour. The
functional $\Sigma[n]$ for the t345 system is the same as for the
hard-core model plus the contribution coming from the attractive
part of the potential:
\begin{eqnarray}
\fl \Sigma[n]=\Sigma^{{\rm (t)}}[n]
-\left( \rho_{\rm v}\sum_nz_n\beta v_ng_0(n,\rho_{\rm v})+
\frac{\rho_{\rm v}^2}{2}\sum_nz_n\beta v_n
\frac{{\rm d}g_0(n,\rho_{\rm v})}{{\rm d}\rho_{\rm v}}\right)
\nonumber \\
\lo\times\left[ \sum_{\lambda\,\,{\rm odd}}(n_{A,\lambda}+n_{C,\lambda}
-2\rho_{\rm v})+
2\sum_{\lambda\,\,{\rm even}}(n_{B,\lambda}-\rho_{\rm v})\right]
\nonumber \\
\lo+ \frac{1}{2}\sum_{\lambda\,\,{\rm odd}}\left[
n_{A,\lambda}\sum_{j|i\in A,\lambda}n_j\beta\Delta v(|i-j|)+
n_{C,\lambda}\sum_{j|i\in C,\lambda}n_j\beta\Delta v(|i-j|)-
2\rho_{\rm v}^2\sum_nz_n\beta v_ng_0(n,\rho_{\rm v})\right]
\nonumber \\
\lo+ \sum_{\lambda\,\,{\rm even}}\left[ n_{B,\lambda}
\sum_{j|i\in B,\lambda}n_j\beta\Delta v(|i-j|)-
\rho_{\rm v}^2\sum_nz_n\beta v_ng_0(n,\rho_{\rm v})\right]
\label{eq35}
\end{eqnarray}
The minimization of (\ref{eq35}) is carried out along the same
lines as for the reference t model, the only difference being in the
novel density functionality of $\Sigma$, not in the way the weighted
density and its derivatives are calculated from the sublattice densities.

However, we expect a number of oddities to follow from (\ref{eq35})
because of the different functional forms of the solid and fluid
free energies. In particular, the minimization of (\ref{eq35})
cannot produce a density profile which, on the $\lambda>0$ side
of the interface, smoothly drops into the vapour one.
In fact, contrary to the cases examined before, $\Sigma[n]$
does not identically vanish when $n_i$ takes the constant value
$\rho_{\rm v}$ (even larger is the difference, at the triple
point, between $\Omega_{\mu}^{{\rm (s)}}(\rho_{\rm l})$ and
$\Omega_{\mu}^{{\rm (f)}}(\rho_{\rm v})$, meaning that the obvious
prerequisite for observing a genuine surface melting is not met).
This mismatch can be quantified in terms of the difference between
$\rho_{\rm v}$ and the homogeneous solution $\rho_{\infty}$ to
$\Omega_{\mu}^{{\rm (s)}}(\rho_{\infty})= \Omega_{\mu}^{{\rm
(f)}}(\rho_{\rm v})$. At $t=1.14$, {\it i.e.}, just below the
triple-point temperature, we find $\rho_{\rm v}=0.0356$ and
$\rho_{\infty}=0.0288$ (the difference being smaller at a lower
$T$).

A way out of this {\it empasse} could be that of imposing $\rho_{\infty}$
as boundary value for $\lambda\rightarrow +\infty$, while maintaining
the form (\ref{eq35}) for $\Sigma[n]$. Obviously, in order to enforce
this condition, the initial {\it ansatz} must be accordingly modified into:
\begin{eqnarray}
n_{A,\lambda} &=& \rho_{\infty}+\frac{n_A-\rho_{\infty}}{1+\exp(\lambda/l)}\,;
\,\,\,(\lambda\,\,{\rm odd})
\nonumber \\
n_{B,\lambda} &=& \rho_{\infty}+\frac{n_B-\rho_{\infty}}{1+\exp(\lambda/l)}\,;
\,\,\,(\lambda\,\,{\rm even})
\nonumber \\
n_{C,\lambda} &=& \rho_{\infty}+\frac{n_B-\rho_{\infty}}{1+\exp(\lambda/l)}\,.
\,\,\,(\lambda\,\,{\rm odd})
\label{eq36}
\end{eqnarray}
We are perfectly conscious that the solution proposed here just
represents a stratagem for making Eq.\,(\ref{eq35}) suited to
describe also the solid surface. A correct description would
in fact need a unique functional for all phases.

For future reference, we plot in Fig.\,9 (top) the MC outcome for
the $x$-integrated densities of the t345 model in a $60\times 128$
slab with periodic boundary conditions along $x$ and fixed densities
at the $y$ boundary. To be precise, the densities are kept fixed at
the $T=0$ solid and vapour values in the eight layers lying on the
extreme left and right of the picture. The temperature is $t=0.87$,
{\it i.e.}, slightly below the exact triple point, whereas the chemical
potential has been adjusted in order to attain phase coexistence. The
occurrence of surface melting in the t345 model is demonstrated by
the structure of the interface in the central part of the picture,
which is compatible with that of a ``modulated'' liquid which
strongly feels the underlying crystal ordering.

In Fig.\,9 (bottom), we have plotted the density profile across
the solid-vapour interface at $t=1.14$, as calculated through the
minimization of $\Sigma[n]$. From a look at this figure, we see
that there are a few layers, interposed between the solid and
the vapour, where the values of the sublattice densities are
intermediate between those of the coexisting solid and vapour and
close, on average, to that of the incoming liquid ($\simeq 0.121$,
at $t=1.15$). Interestingly, a further evidence (see Fig.\,10)
goes in support of the surface-melting interpretation, namely
the existence of a maximum in $n_{B,\lambda}$ near $\lambda=0$,
and of another, less pronounced, in $n_{C,\lambda}$. These maxima
are neither present in the initial profile (\ref{eq36}) nor occur
in the interface profiles of the t and t3 models. Anyway, the
thickness of the molten layer appears to be strongly underestimated
by our DFT as compared to MC. Moreover, also the comparison with
another DFT theory of surface melting~\cite{Ohnesorge} does actually
lead us to qualify our results as rather poor.

We are aware that the use of {\it ad hoc} boundary conditions
in our DFT treatment of the solid-vapour interface may cast some
doubts on the general significance of the results plotted in
Figs.\,9 (bottom) and 10. Certainly, we are not allowed to draw any
reasonable estimate of the surface tension from the calculation
we have presented, which is quantitatively untenable. Notwithstanding
the crudeness of our method, we nonetheless think that Figs.\,9 (bottom)
and 10 do indeed catch the genuine behaviour of the t345 system.

\section{Conclusions}

In this paper, we have used the lattice DFT method to analyse the phase
behaviour of a 2D lattice-gas model (named t345) which exhibits a solid,
a liquid, and a vapour phase. Particles reside on a triangular lattice:
occupation of nearest- and next-nearest-neighbor sites of a particle is
forbidden, while the pair attraction extends from third to fifth neighbors.

We have built up an accurate solid structure for the purely hard-core
model by working with the WDA of Tarazona, while the remaining part of
the t345 potential has been treated as a mean field. This method is
expected to provide good results both at very low and at very high
temperatures, and to offer a not too bad interpolation in between.

As a matter of fact, our theory passes the crucial test of predicting
the existence of a liquid phase in the t345 model. In particular, the
ratio of the triple to the critical temperature is found to reproduce
the exact value to within 2\%. Another successful result is the
prediction, in agreement with MC simulation, of the disappearance of
the liquid phase when the range of the attraction is reduced to embrace
third neighbors only. The main drawback of the theory is in the estimate
of the freezing as well as of the melting density which, in the worst
case, fall short of the exact values by about 35\%. This inconvenient
should be ascribed, besides to the crudeness of the MFA approach (also
worsened by the low system dimensionality), also to the low quality of
the MSA for the hard-core fluid (the HNCA is not viable since it does
not converge even at moderate densities).

Having produced a qualitatively sound bulk theory, we have moved to a
description of the interface structure in the t345 model. The same
functional built up for the bulk system has been used to describe the
coexistence between the solid and the vapour phases. Actually, the use
of slightly different functional forms for the generalized grand
potentials of the solid and of the vapour forces us to introduce a
spurious boundary condition on the vapour side of the interface. If
we allow for this artifice, we do in fact observe the appearance, just
below the triple-point temperature, of a very thin liquid-like layer in
between the solid and the vapour, which is the sign of the occurrence of
surface melting in the system. However, it should be admitted that this
little evidence is not comparable, as for quality, to {\it e.g.} that
provided by the 3D continuum DFT of Ref.\,\cite{Ohnesorge}.

\newpage
\appendix
\section{The lattice DFT of freezing -- generalities}

In this appendix, we first derive a general expression for the
generalized grand potential of an inhomogeneous lattice-gas system;
this is then used for formulating a lattice DFT of freezing. In
particular, we show how to adapt the WDA of Tarazona~\cite{Tarazona1}
to a lattice problem.

Let us suppose to know the DCF $c^{(2)}_{ij}$ of a lattice system and the
value of its $F^{\rm exc}[n]$ for a given density profile $n_0$. Let
$n_{\lambda i}=n_{0i}+\lambda\Delta n_i$, with $\Delta n_i=n_i-n_{0i}$
and $0\leq\lambda\leq 1$. It then follows from the former of
Eqs.\,(\ref{eq10}) that:
\begin{equation}
\beta F^{\rm exc}[n]=\beta F^{\rm exc}[n_0]-\sum_i\Delta n_i
\int_0^1{\rm d}\lambda\,c^{(1)}_i[n_{\lambda}]\,.
\label{a01}
\end{equation}
The functional $c^{(1)}$ can be similarly obtained, using the second
of Eqs.\,(\ref{eq10}), as an integral of $c^{(2)}$, which eventually
yields the exact formula:
\begin{equation}
\fl \beta F^{\rm exc}[n]=\beta F^{\rm exc}[n_0]-\sum_ic^{(1)}_i[n_0]\Delta n_i
-\sum_{i,j}\Delta n_i\Delta n_j\int_0^1{\rm d}\lambda\,
\int_0^{\lambda}{\rm d}\lambda^{\prime}\,c^{(2)}_{ij}[n_{\lambda^{\prime}}]\,,
\label{a02}
\end{equation}
where the first two terms on the r.h.s. of (\ref{a02}) identically
vanish when choosing $n_0=0$.
Actually, an infinite series of terms is hidden behind the last term of
Eq.\,(\ref{a02}), each containing an order of the DCF as calculated for
$n_0$. In practice, one could stop this infinite regression at the second
order by approximating $c^{(2)}_{ij}[n_{\lambda^{\prime}}]$ with
$c^{(2)}_{ij}[n_0]$, and this gives the so called
Ramakrishnan-Yussouff (RY) theory~\cite{Haymet}.

Using Eq.\,(\ref{a02}), the excess free energy per particle of a
fluid with density $\rho$ can be generally written as:
\begin{equation}
\beta f^{\rm exc}(\rho)=
-\frac{1}{\rho}\int_0^{\rho}{\rm d}\rho^{\prime}\,(\rho-\rho^{\prime})
\tilde{c_2}(0,\rho^{\prime})\,,
\label{a03}
\end{equation}
whereas $\rho f^{\rm exc}(\rho)$ is the excess free energy {\it
per site}. The function $c_2$ is calculated by augmenting the OZ
relation with a closure, that is a further relation between the
total and the direct correlation functions. We also quote the
expression of $c_1$:
\begin{equation}
c_1(\rho)=-\beta f^{\rm exc}(\rho)-\rho\beta f^{{\rm exc}\,\prime}(\rho)\,,
\label{a04}
\end{equation}
from which the chemical potential follows through the Eq.\,(\ref{eq11}):
\begin{equation}
\beta\mu=\ln\frac{\rho}{1-\rho}-c_1(\rho)\,.
\label{a05}
\end{equation}

In order to study the coexistence between the solid and the fluid,
we must require equal values of $T$ and $\mu$ for both phases.
Given Eq.\,(\ref{a05}), the departure
$\Delta\Omega[n]=\Omega_{\mu}[n]-\Omega_{\mu}(\rho)$ of the
generalized grand potential of the solid from that of the fluid
can be written as:
\begin{equation}
\fl \beta\Delta\Omega[n]=
\sum_i\left[ n_i\ln\frac{n_i}{\rho}+(1-n_i)\ln\frac{1-n_i}{1-\rho}\right]
+c_1(\rho)\sum_i(n_i-\rho)
+\beta F^{\rm exc}[n]-N\rho\beta f^{\rm exc}(\rho)\,,
\label{a06}
\end{equation}
being $N$ the total number of lattice sites. Every different choice of
$F^{\rm exc}[n]$ defines a class of (approximate) DFTs. The simplest
choice, yet rarely a quantitatively accurate one, is the RY theory,
which leads to:
\begin{equation}
\fl \beta\Delta\Omega[n]=
\sum_i\left[ n_i\ln\frac{n_i}{\rho}+(1-n_i)\ln\frac{1-n_i}{1-\rho}\right]
-\frac{1}{2}\sum_{i,j}c_2(i-j,\rho)(n_i-\rho)(n_j-\rho)\,.
\label{a07}
\end{equation}
The RY theory already represents a considerable improvement over
the ordinary MFA, which is tantamount to assume $c_2(i-j,\rho)=0$ for
$i-j$ inside the core region, and $c_2(i-j,\rho)=-\beta v(|i-j|)$
outside the core. At variance with the MFA, the RY theory uses a
DCF which is adjusted to fit the homogeneous OZ relation as supplemented
with a proper closure. For instance, in the mean spherical
approximation (MSA), one requires that $g(i-j,\rho)=0$ inside the core,
while still assuming $c_2(i-j,\rho)=-\beta v(|i-j|)$ outside this
region. A further possibility would be the Percus-Yevick
approximation (PYA), which assumes $c_2(i,\rho)=g(i,\rho)[1-\exp(\beta
v(|i|))]$.

A more accurate, non-perturbative expression for $F^{\rm exc}[n]$
is obtained by the so called weighted-density approximation
(WDA)~\cite{Tarazona1,Curtin1}, which amounts to approximating
the exact Eq.\,(\ref{a02}) for $n_0=0$ as
\begin{equation}
F^{\rm exc}[n]\approx F^{\rm exc}_{\rm WDA}[n]\equiv
\sum_in_if^{\rm exc}(\bar{n}_i)\,,
\label{a08}
\end{equation}
where the weighted density $\bar{n}_i$ is implicitly defined by:
\begin{equation}
\bar{n}_i=\sum_jn_jw(i-j,\bar{n}_i)\,.
\label{a09}
\end{equation}
The weighted density is required to be constant, {\it i.e.},
$\bar{n}_i=\rho$, for a homogeneous system of density $\rho$ (hence,
$\tilde{w}(0,\rho)=\sum_jw(i-j,\rho)=1$ for any $i$);
moreover, the weight function $w$ must be such that the DCF of the
fluid be recovered in the homogeneous limit:
\begin{equation}
-\beta\left. \frac{\partial^2F^{\rm exc}_{\rm WDA}}
{\partial n_i\partial n_j}\right| _{n_i=\rho}=c_2(i-j,\rho)\,.
\label{a10}
\end{equation}
With the above requirements, the approximation obtained for
$F^{\rm exc}[n]$ is better than any truncated DCF expansion~\cite{Curtin1}.

Using simple calculus, one can translate
Eq.\,(\ref{a10}) into a differential equation for the Fourier
transform of $w(i,\rho)$:
\begin{equation}
\fl -\frac{1}{\beta}\tilde{c_2}(q,\rho)=2f^{\rm exc\,\prime}(\rho)
\tilde{w}(q,\rho)+\rho f^{\rm exc\,\prime\prime}(\rho)
\tilde{w}^2(q,\rho)
+2\rho f^{\rm exc\,\prime}(\rho)\tilde{w}(q,\rho)
\tilde{w}^{\prime}(q,\rho)\,.
\label{a11}
\end{equation}
Although Eq.\,(\ref{a11}) can be numerically solved
for any $q$ and $\rho$~\cite{Curtin1}, we here adopt the simpler
recursive method of Tarazona~\cite{Tarazona1}, which considers a
series expansion of $\tilde{w}(q,\rho)$ in powers of the
density. If we stop at the second order, all we need to
determine is
\begin{equation}
w(i,\rho)=w_0(i)+\rho w_1(i)+\rho^2 w_2(i)
\label{a12}
\end{equation}
from the knowledge of the lower-order terms in the expansions
\begin{equation}
\fl f^{\rm exc}(\rho)=f_1\rho+f_2\rho^2+f_3\rho^3+\ldots
\,\,\,\,\,{\rm and}\,\,\,\,\,
c_2(i,\rho)=\chi_0(i)+\rho\chi_1(i)+\rho^2\chi_2(i)+\ldots
\label{a13}
\end{equation}
We notice that Eq.\,(\ref{a03}) allows us to express $f_{k+1}$
($k=0,1,2,\ldots$) in terms of $\chi_k$ as
\begin{equation}
\beta f_{k+1}=-\frac{1}{(k+1)(k+2)}\sum_i\chi_k(i)\,.
\label{a14}
\end{equation}
Upon inserting Eqs.\,(\ref{a12}) and (\ref{a13}) into Eq.\,(\ref{a11})
and equating term by term, we eventually obtain the general formulae:
\begin{eqnarray}
\fl w_0(i)=-\frac{\chi_0(i)}{2\beta f_1}\,;
\nonumber \\
\fl \tilde{w_1}(q)=-\frac{\tilde{\chi_1}(q)+
4\beta f_2\tilde{w_0}(q)+2\beta f_2\tilde{w_0}^2(q)}
{2\beta f_1\left( 1+\tilde{w_0}(q)\right) }\,;
\nonumber \\
\fl \tilde{w_2}(q)=-\frac{\tilde{\chi_2}(q)+
6\beta f_3\tilde{w_0}(q)+4\beta f_2\tilde{w_1}(q)+
6\beta f_3\tilde{w_0}^2(q)+
8\beta f_2\tilde{w_0}(q)\tilde{w_1}(q)+
2\beta f_1\tilde{w_1}^2(q)}
{2\beta f_1\left( 1+2\tilde{w_0}(q)\right) }\,.
\nonumber \\
\label{a15}
\end{eqnarray}

Given the weight function, the weighted density is explicitly
determined from Eq.\,(\ref{a09}) in terms of the $n_i$ as:
\begin{equation}
\bar{n}_i=\frac{2\bar{n}_{0i}}{1-\bar{n}_{1i}+
\sqrt{(1-\bar{n}_{1i})^2-4\bar{n}_{0i}\bar{n}_{2i}}}\,,
\label{a16}
\end{equation}
where $\bar{n}_{ki}=\sum_jn_jw_k(i-j)$ for $k=0,1,2$. In practice,
one uses Eq.\,(\ref{a16}) as part of the iterative procedure by
which the DFT minimum principle is implemented numerically (see
details in Appendix B). A practical demonstration of the WDA method
will be given in Appendix C.

\newpage
\section{RY theory for the t3 model}

We hereby describe in detail how to work out a RY DFT for the t3
model. Our first task is to determine the DCF of the homogeneous
system. We have chosen to solve numerically the OZ relation with
the MSA closure. Let $a_x=\hat{x}$ and $a_y=(\sqrt{3}/2)\hat{y}$
be the primitive vectors of the
triangular lattice (hereafter, we assume a unit lattice constant).
Then, the reciprocal-lattice vectors are $b_x=2\pi\,\hat{x}$ and
$b_y=(4\pi\sqrt{3}/3)\hat{y}$.
Any sum over Born-Von Karman vectors~\cite{note} is written,
in the infinite-size limit, as an integral over the first
Brillouin zone:
\begin{eqnarray}
\frac{1}{N}\sum_qf(q) &\rightarrow&
\frac{\sqrt{3}}{2}\int_{\rm BZ}
\frac{{\rm d}^2q}{(2\pi)^2}\,f(q)=\frac{\sqrt{3}}{2}\int_{-\pi}^{\pi}
\frac{{\rm d}q_x}{2\pi}\int_{-2\pi\sqrt{3}/3}^{2\pi\sqrt{3}/3}
\frac{{\rm d}q_y}{2\pi}\,f(q_x,q_y)
\nonumber \\
&=& \int_{-\pi}^{\pi}\frac{{\rm d}q_x}{2\pi}\int_{-\pi}^{\pi}
\frac{{\rm d}q^{\prime}_y}{2\pi}\,f(q_x,\frac{2\sqrt{3}}{3}q^{\prime}_y)\,.
\label{b01}
\end{eqnarray}

For the t3 fluid, the MSA assumes: i) $C(3)=c_2(3)=-\beta v_3$
and $C(n)=0$ for all $n>3$ shells; ii) $h(0)=h(1)=h(2)=-1$. From ii),
three equations are derived for the unknown quantities $C(0),C(1)$,
and $C(2)$. For instance, the first of these is obtained by plugging
the OZ relation (\ref{eq15}) into the expression
$h(0)=(1/N)\sum_q\tilde{h}_q$. After a few
manipulations, we eventually obtain the following set of equations:
\begin{eqnarray}
\fl 2\rho(1-\rho)C(3)=\frac{z_3}{(2\pi)^2}
\int_{-\pi}^{\pi}{\rm d}q_x\int_{-\pi}^{\pi}{\rm d}q_y
\frac{1}{1-z_1f_1(q_x,q_y)-z_2f_2(q_x,q_y)-z_3f_3(q_x,q_y)}\,;
\nonumber \\
\fl -6\rho^2C(3)=\frac{z_3}{(2\pi)^2}
\int_{-\pi}^{\pi}{\rm d}q_x\int_{-\pi}^{\pi}{\rm d}q_y
\frac{f_1(q_x,q_y)}{1-z_1f_1(q_x,q_y)-z_2f_2(q_x,q_y)-z_3f_3(q_x,q_y)}\,;
\nonumber \\
\fl -6\rho^2C(3)=\frac{z_3}{(2\pi)^2}
\int_{-\pi}^{\pi}{\rm d}q_x\int_{-\pi}^{\pi}{\rm d}q_y
\frac{f_2(q_x,q_y)}{1-z_1f_1(q_x,q_y)-z_2f_2(q_x,q_y)-z_3f_3(q_x,q_y)}\,,
\label{b02}
\end{eqnarray}
where $z_1=2\rho C(1)/(1-\rho C(0))$, $z_2=2\rho C(2)/(1-\rho C(0))$, and
$z_3=2\rho C(3)/(1-\rho C(0))$ are auxiliary unknowns. Moreover,
\begin{eqnarray}
f_1(q_x,q_y) &=& \cos q_x+2\cos\left( \frac{1}{2}q_x\right) \cos q_y\,;
\nonumber \\
f_2(q_x,q_y) &=& \cos(2q_y)+2\cos\left( \frac{3}{2}q_x\right) \cos q_y\,;
\nonumber \\
f_3(q_x,q_y) &=& \cos(2q_x)+2\cos q_x\cos(2q_y)\,.
\label{b03}
\end{eqnarray}
For a given $\rho$, Eqs.\,(\ref{b02}) are to be solved
recursively: starting from an estimate of $z_1,z_2$ and $z_3$,
these quantities are gradually adjusted until the r.h.s. of
Eqs.\,(\ref{b02}) becomes equal to the quantity on the respective
l.h.s. with a tolerance smaller than $10^{-8}$.

Once the DCF has been determined, we can use {\it e.g.} the RY
theory to construct the generalized grand potential of the t3
model. We call $A$ the triangular sublattice that is occupied in
the $T=0$ crystal, while $B$ includes the rest of the lattice.
Then, the independent density variables are the two sublattice
occupancies, $n_A$ and $n_B$, while the solid density is
$\rho_{\rm s}=(n_A+3n_B)/4$. The density functional that
is to be minimized reads:
\begin{eqnarray}
\fl \frac{4\beta\Delta\Omega(n_A,n_B)}{N}=
n_A\ln\frac{n_A}{\rho}+(1-n_A)\ln\frac{1-n_A}{1-\rho}+
3\left[ n_B\ln\frac{n_B}{\rho}+(1-n_B)\ln\frac{1-n_B}{1-\rho}\right]
\nonumber \\
\lo- \frac{1}{2}\left[ (c_2(0)+6c_2(3))(n_A-\rho)^2+12(c_2(1)+c_2(2))
(n_A-\rho)(n_B-\rho)\right.
\nonumber \\
\lo+ \left. 3(c_2(0)+4c_2(1)+4c_2(2)+6c_2(3))(n_B-\rho)^2\right] \,,
\label{b04}
\end{eqnarray}
having omitted to indicate the $\rho$ dependence of the $c_2$ values.
If, after minimization, $\Delta\Omega$ happens to be
negative, then the solid phase is stable, otherwise the fluid will
overcome the solid in stability.
As a result, the locus of the $\Delta\Omega$ zeroes allows us to
draw the fluid-solid coexistence line in the $T$-$\rho$ plane.

In order to find out the minimum of $\Delta\Omega$, at least two
different strategies can be pursued whose efficiency turns out,
in fact, to be comparable. The first method is to lay down, starting
from somewhere in the $\{n_A,n_B\}$ space, a fictitious relaxational
(steepest-descent) dynamics, {\it i.e.},
\begin{equation}
n_A(t+\Delta t)=n_A(t)-\Delta t\,
\frac{\partial\Delta\Omega}{\partial n_A}(t)\,
\label{b05}
\end{equation}
and similarly for $n_B$, where $\Delta t$ is a conveniently small number.
In the long run, the sublattice densities eventually stabilize, and this
fact will signal that a minimum of $\Delta\Omega$ has been reached
(note that there is always the possibility to get stuck in a local
minimum; so, one should check the nature of the minimum with
different $\Delta t$ values and initial conditions).

The other method is to solve, by a self-consistent procedure,
the non-linear equations for the densities,
\begin{eqnarray}
\fl n_A^{-1}=1+\frac{1-\rho}{\rho}\exp\left[ -(c_2(0)+6c_2(3))(n_A-\rho)
-6(c_2(1)+c_2(2))(n_B-\rho)\right] \,;
\nonumber \\
\fl n_B^{-1}=1+\frac{1-\rho}{\rho}\exp\left[ -2(c_2(1)+c_2(2))(n_A-\rho)
-(c_2(0)+4c_2(1)+4c_2(2)+6c_2(3))(n_B-\rho)\right] \,.
\nonumber \\
\label{b06}
\end{eqnarray}
In order to reach a better convergence, we have resorted to a mixing scheme:
at the $k$-th step in the iteration, we use the inverse of the r.h.s.
of each Eq.\,(\ref{b06}) to obtain a trial estimate (denoted by a hat)
of the densities at the $k+1$-th step. Then, we assume
$n_A^{(k+1)}=(1-q)n_A^{(k)}+q\hat{n}_A^{(k+1)}$ (and similarly for $n_B$),
where $q$ is a small positive number.

The RY freezing and melting lines of the t3 model are showed in Fig.\,5
as dotted lines in the $\rho$-$T$ plane. Since there is only one minimum
in the fluid generalized grand potential, the t3 system shows, according
this theory, two phases only -- fluid and triangular solid, with a density
gap becoming narrower and narrower with increasing temperatures. In the
same figure, we have marked with arrows the densities of the coexisting
fluid and solid in the t model, namely $\rho_f=0.1495$ and $\rho_s=0.1600$
(see below). In fact, the t model can be viewed as the infinite-temperature
limit of the t3 model.

The MSA equations for the t model can be easily adapted from those
of the t3 model. An important thing to notice is that the iterative
procedure by which the MSA is solved does usually fail to converge
beyond a certain density $\rho_{\rm up}$ which, for the t model, is
slightly above $0.21$. This is a well-known problem in the field
of integral equations of classical fluids which, however, is not
particularly dangerous in view of the fact that the fluid phase
loses its stability against the solid well below $\rho_{\rm up}$.
This notwithstanding, we might have the need to extend, as required
by the forthcoming WDA of Appendix C, the definition of
$f^{\rm exc}(\rho)$ well beyond $\rho_{\rm up}$ (and even beyond 0.25).
To this end, since the only obvious constraint to fulfill is
regularity, the possible solutions are many. Following the
proposal advanced in Ref.\,\cite{Baus} for hard disks, we could
assume, for instance, the (metastable-) fluid pressure to be
exactly given, beyond $\rho=0.21$, as:
\begin{equation}
\frac{\beta P}{\rho}=\frac{1+a^{\prime}\eta+b^{\prime}\eta^2+
c^{\prime}\eta^3+d^{\prime}\eta^4}{(1-\eta)^2}\,,
\label{b07}
\end{equation}
where $\eta=(2\pi\sqrt{3}/3)\rho\equiv\alpha\rho$ is the packing
fraction (corresponding to a hard-core diameter of 2), while
$a^{\prime},b^{\prime},c^{\prime}$, and $d^{\prime}$ are free
parameters. The excess free energy will follow from
\begin{equation}
\beta f^{\rm exc}(\rho)=\int_0^{\rho}\left( \frac{\beta P(t)}{t}-1\right)
\frac{{\rm d}t}{t}\,,
\label{b08}
\end{equation}
which, through the density derivative of Eq.\,(\ref{a05}), is easily
proved to be equivalent to Eq.\,(\ref{a03}). Upon inserting
Eq.\,(\ref{b07}) into
Eq.\,(\ref{b08}), we eventually arrive at the following analytic form:
\begin{equation}
\beta f^{\rm exc}(\rho)=a\rho+b\rho^2+\frac{c\rho}{1-\alpha\rho}+
d\ln(1-\alpha\rho)\,,
\label{b09}
\end{equation}
with other parameters $a,b,c$, and $d$. The latter are fixed by requiring
a smooth behaviour at $\rho=0.21$.

The problem with the above extrapolation (called E1) is that
(\ref{b09}) blows up to infinity for $\rho=1/\alpha\simeq 0.276$.
This could be a serious inconvenient if one needs to calculate
$f^{\rm exc}(\rho)$ beyond $1/\alpha$. In this case, we resort to
a simpler extrapolation (called E2), which merely expresses
$f^{\rm exc}(\rho)$ as a fourth-order polynomial beyond $\rho=0.21$.

\newpage
\section{WDA for the t model}

In the present appendix, we show how to build up a WDA theory for
the t model.

Upon inserting (\ref{a08}) into (\ref{a06}), and specializing to
the t model, we readily obtain:
\begin{eqnarray}
\fl \frac{4\beta\Delta\Omega(n_A,n_B)}{N}
=n_A\ln\frac{n_A}{\rho}+(1-n_A)\ln\frac{1-n_A}{1-\rho}+
3\left[ n_B\ln\frac{n_B}{\rho}+(1-n_B)\ln\frac{1-n_B}{1-\rho}\right]
\nonumber \\
\lo+ c_1(\rho)(n_A+3n_B-4\rho)
+n_A\beta f^{\rm exc}(\bar{n}_A)+3n_B\beta f^{\rm exc}(\bar{n}_B)-
4\rho\beta f^{\rm exc}(\rho)\,.
\nonumber \\
\label{c01}
\end{eqnarray}
If we impose the vanishing of the partial derivatives of (\ref{c01}),
we get the equations for $n_A$ and $n_B$:
\begin{eqnarray}
\fl n_A^{-1}=1+\frac{1-\rho}{\rho}\exp\left[
c_1(\rho)+\beta f^{\rm exc}(\bar{n}_A)+n_A\beta
f^{\rm exc\,\prime}(\bar{n}_A)\frac{\partial\bar{n}_A}{\partial n_A}+3n_B\beta
f^{\rm exc\,\prime}(\bar{n}_B)\frac{\partial\bar{n}_B}{\partial n_A}\right] \,;
\nonumber \\
\fl n_B^{-1}=1+\frac{1-\rho}{\rho}\exp\left[
c_1(\rho)+\beta f^{\rm exc}(\bar{n}_B)+\frac{1}{3}n_A\beta
f^{\rm exc\,\prime}(\bar{n}_A)\frac{\partial\bar{n}_A}{\partial n_B}+n_B\beta
f^{\rm exc\,\prime}(\bar{n}_B)\frac{\partial\bar{n}_B}{\partial n_B}\right] \,.
\nonumber \\
\label{c02}
\end{eqnarray}
In the above equations, the weighted densities $\bar{n}_A$ and $\bar{n}_B$
are calculated from Eq.\,(\ref{a16}). As for their density derivatives,
it follows from the original definition (\ref{a09}) that
\begin{equation}
\frac{\partial\bar{n}_A}{\partial n_A}=
\left( 1-\bar{n}_{1A}-2\bar{n}_{2A}\bar{n}_A\right) ^{-1}
\left( \frac{\partial\bar{n}_{0A}}{\partial n_A}+
\bar{n}_A\frac{\partial\bar{n}_{1A}}{\partial n_A}+
\bar{n}_A^2\frac{\partial\bar{n}_{2A}}{\partial n_A}\right) \,,
\label{c03}
\end{equation}
and similarly for other derivatives. In Eq.\,(\ref{c03}),
$\bar{n}_{kA}=\sum_jn_jw_k(i-j)$, with $i\in A$ and $k=0,1,2$.
We thus have, for instance,
\begin{eqnarray}
\fl \frac{\partial\bar{n}_{kA}}{\partial n_A}=\sum_{j\in A|i\in A}w_k(i-j)
\nonumber \\
\lo=w_k(0)+6w_k(3)+6w_k(6)+6w_k(8)+12w_k(13)+6w_k(15)+6w_k(19)+\ldots\,,
\nonumber \\
\label{c04}
\end{eqnarray}
where we have used the shell number (rather than the distance) as
argument for $w_k$. Obviously, in order to make the whole
procedure computationally feasible, the sum in Eq.\,(\ref{c04})
(and any other sum of the same kind) must be truncated at a
certain distance, and we have chosen to stop summing beyond the
distance (equal to 7) of the 20th neighbors. This is not a
problem, however, since the $w_k$ functions rapidly drop to zero
when increasing the distance from the reference site.

We remark that a novel feature emerges in the behaviour of $w_k(i)$
right when we reach the distance of the 20th neighbors, which is not
observed at the smaller distances. Two different groups of such
neighbors are, in fact, to be distinguished: 6 of them are
symmetry related, as are the other 12. But the value of $w_k$ for
a site of the first group is {\it different} from that calculated
for a neighbor of the second group (hence we have a $w_k(20a)$ and
a $w_k(20b)$). We should wait until the 33th-neighbor shell (at a
distance of $\sqrt{91}$) to observe this feature repeated again.
Hence, notwithstanding the potential shows radial
symmetry, $w_k$ is {\it not} spherically symmetric and this is the
reason why, on a lattice, particular care must always be paid to
distinguish translational from spherical symmetry, although the
exceptions to radial symmetry are, in a sense, rare~\cite{Prestipino2}.
Note that the reduced PDF behaves similarly to $w_k$, {\it i.e.},
$g(i-j,\rho)$ is not spherically symmetric either.

\newpage
\begin{center}
{\bf Figure Captions} \vspace{5mm}
\end{center}
\begin{description}
\item[{\bf Fig.\,1 :}] The t model (MSA+WDA). Left panel: total
density ($\rho$ in the fluid phase, $n_{\rm s}$ in the solid phase)
vs. reduced chemical potential. The sublattice densities, $n_A$ and
$n_B$ (dotted lines), are also plotted in the solid region.
Right panel, solid region: the weighted densities, $\bar{n}_A$ and
$\bar{n}_B$, vs. reduced chemical potential (we have used E1 for
extending the definition of $f^{\rm exc}(\rho)$ beyond $\rho=0.21$).

\item[{\bf Fig.\,2 :}] Phase diagram of the t345 model, using the
t model (MSA+WDA) as a reference. Two distinct approximations for
the perturbation part are compared through the respective phase
diagrams: Eq.\,(\ref{eq21}) ($\bigcirc$) vs. the MFA ($\times$).
The freezing and melting lines are constructed through the use of
E1 at high $T$ and of E2 at low $T$. For comparison, we show as
asterisks some MC data points for a $48\times 48$ lattice (MC
averages are taken over $5\cdot 10^5$ equilibrium sweeps; the
errors affecting these points are of the same size as the symbols).
The lines connecting the points are just a guide for the eye.
The arrows pointing downwards mark the densities of the
coexisting fluid and solid in the t model, as drawn from MSA+WDA.
The other arrows mark the MC values for the same quantities.

\item[{\bf Fig.\,3 :}] The t345 model (MSA+WDA+perturbation). The
picture shows the $\beta\mu$-evolution of the overall density --
$\rho$ for the fluid and $n_{\rm s}=(n_A+3n_B)/4$ for the solid --
along the isotherm $t=1.2$. In the solid region, the real and
weighted densities are separately plotted for the two sublattices
$A$ and $B$ as dotted lines (note that $n_B$ and $\bar{n}_A$ are
almost indistinguishable). The dotted vertical lines mark the points
where the phase transitions occur.

\item[{\bf Fig.\,4 :}] The t345 model (MSA+WDA+perturbation). The
DFT phase diagram of the t345 model ($\bigcirc$) as it appears on
the $T$-$\mu$ plane. The solid-fluid coexistence line is of the E1
type at high $T$, and of the E2 type at lower $T$ values. For
comparison, we have also reported as asterisks the MC data for a
$48\times 48$ lattice. Straight lines are drawn through the symbols
as a guide for the eye. The inset shows a zoom on the triple-point
region.

\item[{\bf Fig.\,5 :}] Phase diagram of the t3 model. Two distinct
DFTs are contrasted through the respective t3 phase diagrams: one
theory uses the t model (MSA+WDA) as a reference and the attractive
interaction as a perturbation ($\bigcirc$; the freezing and melting
lines are constructed through the use of E1 at high $T$ and of E2
at low $T$); the other theory is MSA+RY ($\times$). MC data for a
$48\times 48$ lattice and $2\cdot 10^5$ equilibrium sweeps are shown
as asterisks.
The straight lines between the points are plotted as a guide for
the eye. The densities of the coexisting fluid and solid in the t
model are marked as downward-pointing arrows (long and short arrows
are for the MSA+WDA and the RY theory, respectively). The other arrows
mark the MC estimates.

\item[{\bf Fig.\,6 :}] The homogeneous t model. Two distinct closures
of the OZ relation are compared through the profile of the reduced PDF
at $\rho=0.1$: MSA ($\bigcirc$ and dashed line) and HNCA ($\triangle$
and dotted line). At the distances $r_{20}=7$ of the 20th neighbors
and $r_{33}=\sqrt{91}$ of the 33th neighbors, two symbols are reported
for each curve (see the discussion following Eq.\,(\ref{c04})).
The full dots are the MC data points for a $48\times 48$ sample at
$\beta\mu=-0.32$ (here, the average density is
$\left< c_i\right>=0.09995(1)$ over $5\cdot 10^5$ equilibrium sweeps).
Inset: a magnification of the
large-distance region. It distinctly appears that the MSA PDF
is of an overall better quality than the HNCA one.

\item[{\bf Fig.\,7 :}] The t345 model (MSA+WDA+perturbation).
The figure shows the density profile across the interface between
the coexisting liquid and vapour phases at $t=1.15$. The starting
point of the functional minimization is an exp profile (dotted line);
The open dots (which, for clarity, are joined by a continuous line)
are the final outcome of the optimization. It turns out that the
difference between the two curves is very small.

\item[{\bf Fig.\,8 :}] Top: the t3 model (MSA+RY), density profile
across the solid-fluid interface at $t=1.8036$ (here,
$\rho_{\rm f}=0.1000$ and $\rho_{\rm s}=0.1695$).
Bottom: solid-fluid interface in the t model (MSA+WDA). In both
panels, the optimal exp profile (dotted line) is contrasted with
the outcome of an unconstrained $\Sigma[n]$ minimization (open
dots and continuous line).

\item[{\bf Fig.\,9 :}] Top: MC density profile across the solid-vapour
interface of a t345 lattice system at $t=0.87$, {\it i.e.}, just below
the triple-point temperature. To reach coexistence, the chemical potential
is set equal to $\mu=-4.479\,V$. For a simulation box of $60\times 128$,
as many as $2\cdot 10^6$ equilibrium sweeps were produced. The dotted
line marks the average density over couples of adjacent layers. Near
the centre of the picture, the maximum in the interstitial density
(which is the bottom of the modulation) is the sign of a liquid-like
behaviour.
Bottom: DFT results for the t345 model (MSA+WDA+perturbation).
The density profile across the solid-vapour interface is shown at
$t=1.14$: optimal exp profile (dotted line) vs. unconstrained
$\Sigma[n]$ minimization (open dots and continuous line).

\item[{\bf Fig.\,10 :}] The t345 model (MSA+WDA+perturbation).
The profile of $n_{B,\lambda}$ and $n_{C,\lambda}$ vs. $\lambda$
for the same solid-vapour interface being represented in Fig.\,9
(bottom). The unconstrained minimum-$\Sigma[n]$ profile (open
dots and continuous lines) is compared with the best exp
{\it ansatz} (dotted line). The maxima near $\lambda=0$ are
a sign of the local onset of a liquid-like behaviour.
\end{description}

\begin{thebibliography}{99}
\bibitem{note2} It is clear that what we are referring to here as a ``solid''
is just one of the four symmetry-related ergodic components which
characterize the very dense system, {\it i.e.}, the
broken-symmetry phase which is singled out after choosing a
condensation seed. Otherwise, the solid phase itself would be
homogeneous.
\end{thebibliography}
\end{document}